\documentclass[12pt]{iopart}
\usepackage{iopams}
\usepackage{afterpage}
\usepackage{graphicx}
\usepackage{amssymb}

\newcommand{\bq}{\begin{equation}}
\newcommand{\ee}{\end{equation}}
\newcommand{\fr}[2]{\frac{#1}{#2}}
\newcommand{\eps}{\varepsilon}

\begin{document}

\title[Level-occupation switching and phase anomalies...]{
Level-occupation switching of the Quantum Dot, and phase anomalies
in mesoscopic interferometry}

\author{P.G.Silvestrov$^{1}$ and Y.Imry$^{2}$}
\address{$^{1}$ Theoretische Physik III, Ruhr-Universit{\"a}t Bochum,
44780 Bochum, Germany\\
$^{2}$ Weizmann Institute of Science, Rehovot 76100, Israel}
\begin{abstract}
For a variety of quantum dots, the widths of different
single-particle levels may naturally differ by orders of magnitude.
In particular, the width of one strongly coupled level may be larger
than the spacing between other, very narrow, levels. We found that
in this
case many consecutive Coulomb blockade peaks 
are due to occupation of the same broad level. Between the peaks
the electron jumps from this level to one of the narrow levels and
the transmission through the dot at the next resonance essentially
repeats that at the previous one. This offers a natural
explanation of the salient features of the behavior of the
transmission phase in an interferometer with a QD. The theory of
this effect will be reviewed with special emphasis on the role of
the interactions. New results on the dot-charging measurements and
the fine structure of occupation switchings will be presented,
accompanied by the unified description of the whole series of CB
peaks caused by a single broad level. We then discuss the case
where the system approaches the Kondo regime.
\end{abstract}

\maketitle

\section{Introduction}

The  challenging results of  the experiment~\cite{Heiblum2} which
employed the double slit interference scheme in order to determine
the phase~\cite{endnote}
of the wave transmitted
through a quantum dot~(QD) in the Coulomb-blockade~(CB)~\cite{CB}
regime is discussed in several articles of this volume. The
unexpected result of this experiment was a repeated fast jump of
the phase by $-\pi$ between the resonances near the minimum of the
transmitted current. [The phase at each CB peak increased smoothly
by $\pi$ in accordance with the Breit-Wigner picture.] In this
paper we will first review the mechanism intended to explain these
experimental findings suggested in our paper Ref.~\cite{SIPRL} and
then discuss new features and results. The essence of our
suggestion in Ref.~\cite{SIPRL} was the observation that if a QD
happens to support a single level extremely well coupled to the
leads, the conductance for a series of charging resonances will be
governed by the transport through (population of) this level.
Other, narrow, levels are occupied, and the broad level is
repeatedly depopulated via (almost) abrupt occupation jumps in the
CB valley. This interlevel electron-exchanging mechanism, which
has later been dubbed "population switching" (although the notion
"level-occupation switching" may better reflect the physical
situation), can lead to a satisfactory explanation of the
experiment. The first suggestion of occupation switching as a
possible explanation for the phase lapses and peak correlations
(see below) was given in Ref.~\cite{Hackenbroich} and followed up
in Ref.~\cite{Gefen}. However, that mechanism was due to an
assumed special structure of the dot, not to electron correlations
as in the present work.

Another important result of the experiment~\cite{Heiblum2}, which
also disagrees with a number of models, is the similar shape and
height of consecutive CB peaks. This effect, which is often not
emphasized enough, is again naturally explained within our picture.

Experiments of the Weizmann
group~\cite{Heiblum2,Heiblum1,Yang1,Yang2,Michal} measuring the
transmission phase in QD-s have stimulated a wide theoretical
interest, which is (partly) reflected by Refs.~\cite{Hackenbroich,
Gefen, EAIH, Weidenmuller, butt, Deo, Fano, Fano1, KarrFano,
KarrNJP, Oreg, Baltin, SIPRB2002, SIKondo, vonDelftPRB05, Lee06,
GG, Silva02, Hersh, Oreg1, Hacken, Delft, Hof}. In particular, the
phase behavior observed in Ref.~\cite{Heiblum2}: an increase by
$\pi$ at the resonance accompanied by a sharp $-\pi$ jump in the
CB valley, is in evident contradiction with what one would expect
had the transmission of the current proceeded via consecutive
levels in a 1-dimensional quantum well. In a two--dimensional QD
the phase drops associated with the nodes of the transmission
amplitude arise already within the single--particle
picture~\cite{Deo,Fano,Fano1}, related to the Fano effect.
However, in order to have a sequence of such events one should
consider a QD of a very special shape or level structure, which is
not expected in the generic case~\cite{butt}. The model of
Ref.~\cite{Oreg} also does not allow to explain the series of
drops, especially at very low temperatures. The mechanism of
Refs.~\cite{Hackenbroich,Gefen}, as we already mentioned, makes
nontrivial assumptions on the geometry of the QD and the way it
changes under the change of plunger gate voltage. A generic
mechanism suggested in Ref.~\cite{Baltin} may indeed lead to the
correlations in transmission at many consecutive valleys, but the
predicted phase behavior differs from what has been seen
experimentally.

The mechanism for charging of the QD, which we will discuss in
detail in the next section includes two important ingredients. One
is the necessity of the existence of levels anomalously coupled to
the leads in sufficiently large irregular QD-s. The mechanism
suggested for this purpose in our paper Ref.~\cite{SIPRL} will be
reviewed in Sec.~2.2. Further developments of this and related
mechanisms may be found in Refs.~\cite{butt, Hacken, vO}. In all
these cases the broadening of the single particle level does not
require the electron interaction in the QD.

The second ingredient is the interaction-induced occupation
switching, which we describe in section~\ref{sec2.1}. Further
developments or applications of this effect, also discussing
applications not related to the transmission phase problem, may be
found in Refs.~\cite{vonDelftPRB05, sense, Lee06, KG, GG, Dagotto,
SIPRB2007}. Finally, recent papers ~\cite{KarrFano,KarrNJP}
investigated, invoking the Fano effect and the dynamically
generated level broadening~\cite{vO}, both the universal regime,
where phase lapses generically occur, and the mesoscopic regime,
where the phase behavior is irregular, as in Ref.~\cite{Michal}.
The crossover between these two regimes as a function of a single
parameter, was also discussed.

Our main effort in the first parts of this paper will be to explain
clearly the basic  physics of the effect of level-occupation
switching in the simplest cases, precisely defining the effect of
the interaction. Effects of the electron spin~\cite{SIPRB2002} will
be discussed in Sec.~3, while new results on the details of
occupation switching will be treated in Sec.~4. The case where the
QD approaches the Kondo regime~\cite{SIKondo} will be considered in
Sec.~5 in relation with the experiment of Ref.~\cite{Yang1}. We
conclude with some remarks on where the theory stands vis a vis the
explanation of the experiments.

\section{A model for the level-occupation switching}

A useful theoretical model for the description of charging effects
in QD-s is the tunneling Hamiltonian in the constant interaction
($U_{CB}$) approximation (see e.g.~\cite{CB})
 \begin{eqnarray}\label{Ham}
 H&=&\sum_i \eps_i c^+_i c_i + U_{CB}\sum_{i<j} c^+_i c_i c^+_j
c_j\\
 &+&   \sum_k\eps(k)^L a^{L+}_k a^L_k + \sum_{k,j}[t^L_j c^+_j a_k^L
+{\rm H.c.}] + L\leftrightarrow R \ . \nonumber
 \end{eqnarray}
Here $c(c^+)$ and $a(a^+)$ are the annihilation(creation)
operators for electrons in the dot and in the lead and $\eps_i,
\eps(k)$ are the single-particle energies. We do not introduce the
$k$ dependence of the tunneling matrix elements $t_j$. Summation
over spin orientations is also easily included, but we first treat
the spinless case (which is experimentally realizable by applying
a strong enough magnetic field to the dot. To avoid the orbital
effects of the field, it may be applied parallel to the plane of
the dot). Under the assumption of capacitive coupling to the gate,
the levels in the dot flow uniformly with the voltage
 \bq\label{1}
\eps_i=\eps_i{(V_g=0)} -V_g \ .
 \ee
The energy of an electron in the wire with momentum $k$ is given
by
 \bq\label{11}
\eps(k)=\fr{k^2}{2m}-E_F \ .
 \ee

Although the experiment~\cite{Heiblum2} was clearly done in the CB
regime, the widths of the resonances turned out to have been
anomalously large, only few times smaller than the charging energy
and larger than the dot's level spacing. Also, the widths and
heights of all observed resonances are very similar. These
surprising features of the results of Ref.~\cite{Heiblum2}, which
have not attracted so wide an attention as the phase jumps,
motivate us to consider the model in which the QD supports a
single very broad (well coupled to the leads) level, which we
describe in the next subsection. Widths of all other levels will
be assumed to be small and may be set to zero in the leading
approximation.

It is generally believed that the CB is observed only if the widths
of resonances are small compared to the single-particle level
spacing in the dot $\Delta$ \cite{CBdest}. This condition assumes
that the couplings of all levels to the leads are of the same order
of magnitude. However, as we will show below in section~2.2, for
many situations the widths of the resonances may vary by orders of
magnitude. In this case it does not make sense to compare the width
of few broad resonances with the level spacing, determined by the
majority of narrow, practically decoupled, levels. [Even the case of
a single level being wider than the charging energy $U_{CB}$ may be
considered within our approach~\cite{sense}.]

\subsection{The case of a single broad level\label{sec2.1}}

Now we turn to the many-particle effects arising for the
Hamiltonian~(\ref{Ham}) in the case of only one ($N$-th) level in
the dot coupled strongly to the wires, $|t^{L,R}_N|\gg
|t^{L,R}_j|$, $j\ne N$. If the width of this level is larger than
the single-particle level spacing $\Delta$, a very nontrivial
regime of charging of the QD may be described by means of
second-order perturbation theory estimates (compare with
Ref.~\cite{Haldane}). This simple treatment produces all the
important features of the problem (a similar approach was used
recently in Ref.~\cite{Kaminski} for the calculation of CB peaks
positions).

The width of the well coupled level is given by
 \bq\label{2}
\Gamma\equiv \Gamma_N = \Gamma^L_N+\Gamma^R_N=2\pi \sum_{i=L,R}
|t^i_N|^2 \fr{dn}{d\eps} \gg \Delta \ .
 \ee
The widths of the other levels are taken to be much smaller than
the level spacing and may be neglected in the first approximation.
The charging energy is still  very large $U_{CB}\gg\Gamma$. We
shall show that transmission of a current at about
$(\Gamma/\Delta)ln(U_{CB}/\Gamma)$ consecutive CB peaks will
proceed through one and the same level $\eps_N$. Let the levels
with $i\le 0$ in the QD be occupied. Our aim is to find the total
energy of the true ground state of the dot at different values of
$V_g$. Without loss of generality we may assume that the summation
over $i$ in Eq.~(\ref{Ham}) goes only over $i>0$. (Thus we
subtract from the total energy the trivial constant corresponding
to selfinteraction of electrons with $i\le 0$. Coulomb interaction
between electrons at the levels with $i\le 0$ and $i>0$ is
included into $\eps_{i>0}$.) We also subtract from the total
energy the trivial energy of electron gas in the leads $\sum
\eps(k) \langle a^{+}_k a_k \rangle$.

\begin{figure}
    \centering
\includegraphics[width=0.8\textwidth,clip]{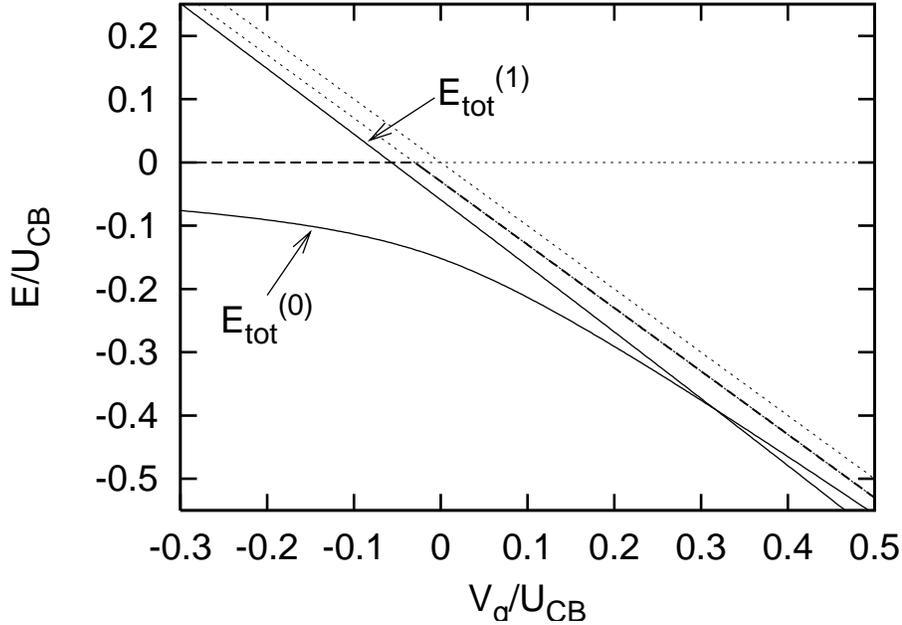}
\caption{The evolution of the ground state energy of the QD with a
single level strongly coupled to the leads, described in the text.
Dashed lines show the energies for different states of the dot if
the coupling to the leads would be switched off completely. The thin
horizontal line is the energy of the empty dot, $E=0$. Tilted dashed
lines show the energy of the decoupled QD with the electron in state
$j$, $E=\eps_j$ (lower line), and with the electron in state $N$,
$E=\eps_N$ (upper line). The ground state of the decoupled QD
follows the thick dashed line. After the coupling of the level $N$
to the leads is switched on the energies of the states having
initially no electrons in the dot, $E_{tot}^{(0)}$, and one electron
in the state $j$, $E_{tot}^{(1)}$, are renormalized, as described by
Eqs.~(\ref{E0},\ref{5},\ref{4},\ref{6}). The crossing of the two
solid lines in the right lower part of the figure shows the
occupation switching, Eq.~(\ref{8}). } \label{Atan}
\end{figure}

In this section we consider spinless electrons. We use a superscript
$(n)$ in the total energy of the many-electron ground state to
denote the number of electrons, $n$, in the narrow levels (with
$i>0$) in the QD. Thus $E^{(0)}_{tot}$  and $E^{(1)}_{tot}$ stand
for zero and one electron in the narrow levels, respectively. We
start with $(0)$ and with the case of large positive $\eps_N(V_g)\gg
\Gamma$. The only nontrivial contribution to the total energy is
given by the second order correction where an electron hops from the
lead to the level $N$ and back,
 \bq\label{E0}
E^{(0)}_{tot}=\sum_{i=L,R}\int_{-E_F}^0
\fr{|t^i_N|^2}{\eps-\eps_N} \fr{dn^i}{d\eps} d\eps = \fr{-
\Gamma}{2\pi} \ln\left( \fr{4E_F}{\eps_N} \right) .
 \ee
To calculate the integral here we use that
$(dn/d\eps)d\eps=(L/\pi)dk$, where $L$ is the length of the wire and
the energy and momentum are related by Eq.~(\ref{11}). In
Eq.~(\ref{E0}) and everywhere below, $\eps_N,\eps_j$ are functions
of $V_g$~(\ref{1}). When with the increasing voltage $V_g$ the
decreasing energy level $\eps_N$ crosses the region
$|\eps_N|\sim\Gamma$ the level $N$ becomes occupied (occupation of
the level $j$ is energetically unfavorable, as we will see below,
even though $\eps_j<\eps_N$).

The total energy of the system (with all the subtractions described
above) may be calculated similarly to Eq.~(\ref{E0}) for $\eps_N \ll
-\Gamma$ (the broad level being well below the Fermi energy). The
two contributions to $E^{(0)}_{tot}$ are now the energy of the
electron in the dot $\eps_N$ (which was taken where from the state
$\eps(k_F)=0$ in the lead~(\ref{11})) and the second order
correction describing the electron hopping from the level $N$ to the
leads. This gives
 \bq\label{5}
E^{(0)}_{tot}= \eps_N -\fr{\Gamma}{2\pi} \ln\left(
\fr{4E_F}{|\eps_N|} \right) \, .
 \ee

Alternatively, Eqs.~(\ref{E0}) and (\ref{5}) may be derived without
explicit reference to many-body perturbation theory. To this end we
first perform a unitary rotation of the lead states such that only a
single effective lead defined by the annihilation operator
$a_k=(t^La^L_k+t^Ra^R_k)/\sqrt{|t^L|^2+|t^R|^2}$ remains coupled to
the dot. In the absence of the QD level the levels in this lead are
quantized via $k_n=n\pi/L, n=1,2,\cdots,\infty$. Adding a discrete
QD level to this dense grid of the lead levels results in a slight
shift of the new levels with respect to the old ones (see e.g. the
book~\cite{Bohr}). The levels with $\eps(k)<\eps_N$ are shifted
downwards, while the ones with $\eps(k)>\eps_N$ are shifted upwards.
The individual shift of any level is smaller than the level spacing
in the lead ($\propto 1/L$), but the accumulated effect of the
shifts of many ($\propto L$) levels lead to a visible $\sim\Gamma$
correction to the total energy. The sum of the downward shifts of
all (full) levels with $\eps(k)<0$ in case of $\eps_N>0$
($\eps_N\gg\Gamma$) gives Eq.~(\ref{E0}). The second term in
Eq.~(\ref{5}) (the case of $\eps_N<0$) contains two contributions: a
negative one due to the shifts of levels with $\eps(k)<\eps_N$, and
a positive due to the shifts of the levels with $\eps_N<\eps(k)<0$.
A perturbative calculation of this correction becomes possible
because of the cancellation of contributions from regions
$2\eps_N<\eps(k)<\eps_N$ and $\eps_N<\eps(k)<0$.

Finally, a precise treatment for a single state interacting with a
continuum for spinless electrons yields~\cite{SIPRL}:
 \bq\label{4} E^{(0)}_{tot} =
\fr{-\Gamma}{4\pi} \left[
\ln\left(\fr{16E_F^2}{\eps_N^2+\Gamma^2/4}\right)
 +2 \right]
 + \fr{\eps_N}{\pi}
\cot^{-1}  \fr{2\eps_N}{\Gamma}  ,
 \ee
which coincides with the Eqs.~(\ref{E0},\ref{5}) at
$|\eps_N|\gg\Gamma$ and gives the correct interpolation between
them.

Let us now consider the branch where the level $j$ is occupied
(which we label by the superscript $(1)$). The energy of the
electron in the QD is $\eps_j$. However, the energy of the empty
level $N$ is now increased by $U_{CB}$, {\it due to the
interaction}. Adding one more electron via the hopping $t_N$ from
the Fermi level in the lead, to that level, {\em now costs
$\eps_N+U_{CB}$}. The ensuing reduction of the downward shift of the
level $E^{(1)}_{tot}$ is of crucial importance. This is precisely
where the effect of the interaction is felt. The analog of Eq.
(\ref{E0}) for $V_g<U_{CB}$ now reads
 \bq\label{6}
E^{(1)}_{tot}= \eps_j -\fr{\Gamma}{2\pi} \ln\left(
\fr{4E_F}{\eps_N+U_{CB}} \right) \, .
 \ee

Without the interaction $U_{CB}$, the two second-order correction
terms on the RHS's of Eqs.~(\ref{5}) and~(\ref{6}) are equal. This
explains why these corrections to the total energy of the system
do not affect the order of level populations in the noninteracting
QD. For the case considered, $\eps_j < \eps_N$, level $j$ will be
the first to be filled when its energy crosses the Fermi level in
the leads, $\eps_j(V_g)=0$, as is shown by a thick dashed line in
Fig.~1.

With the interaction $U_{CB}$, after the electron is added to the
dot the second order corrections in Eqs.~(\ref{5}) and (\ref{6})
start to depend on which of the orbitals $N$ or $j$ is occupied by
this electron. In a large part of the CB valley the energy gain
because of the increased second order correction overcomes the
loss, $\eps_N-\eps_j$, caused by the raising the electron from the
level $j$ to the level $N$. The true ground state of the system
here will be $E^{(0)}_{tot}$~(\ref{5}). The two functions
$E^{(0)}_{tot}(V_g)$ and $E^{(1)}_{tot}(V_g)$ cross at
 \bq\label{8}
\eps_N\approx -\fr{U_{CB}}{\exp\{ 2\pi(\eps_N -\eps_j)/\Gamma\}
+1} \ ,
 \ee
and the ground state jumps onto the branch $E^{(1)}_{tot}$. Here the
"active" electron inside the dot is transferred from the broad level
$N$ to the narrow one $j$. The energy of the current-transmitting
{\it virtual} state $N$ becomes  positive again, $\eps_N+U_{CB}>0$.
Thus, the phase of the transmission amplitude has returned to what
it was before the process of filling of level $N$ and the subsequent
sharp jump into the state where level $j$ is filled. It is the
latter jump which provides the sharp drop by $\pi$ of the
transmission  phase, following its gradual increase by $\pi$ through
the broad resonance. Fig. \ref{Atan} illustrates the behaviors of
the bare levels and the renormalized ones, including the crossing of
the latter at the switching point.

For the case considered so far, where the width of the narrow
level vanishes, the switching transition is infinitely sharp.
However when the coupling of level $j$ is switched on,
second-order processes involving jumps from level $j$ to the lead
and back from the lead to level $N$ (or vice versa) will create a
matrix element connecting these two states. This makes the
crossing of the energies $E^{(0)}_{tot}(V_g)$ and
$E^{(1)}_{tot}(V_g)$, considered above,  an "avoided" one, and
provides a finite width for the switching transition.

Even when the switching transition is broadened, the phase jump
should still be abrupt (has zero width) in case of spinless
electrons, as was proven in Ref.~\cite{SIPRB2002}. Fast phase
drops in the valley are associated with nodes of the amplitude,
$A\sim \eps-\eps_{0}$, where $\eps_0$ is  close to the real axis
(small ${\rm Im}\eps_0$). For spinless electrons,  the linear
conductance (even through the interacting QD, since we are
interested in the elastic processes) is determined by a single
complex amplitude, $G\propto |A|^2$, which is the amplitude of
transmitted electron wave $Ae^{ikx}$ in e.g. the right lead for
unit amplitude of the incoming wave, $1\times e^{ikx}$, in the
left lead. [In case of spin the linear conductance would be given
by a sum of the squared moduli of several complex amplitudes
describing the scattering for different orientations of the
electron spin and the spin of the QD, including spin-flip
amplitudes, while the Aharonov-Bohm phase will be determined by
the phase of the sum of non-spin-flip amplitudes, as discussed in
Sec.~5 below]. The vanishing of the transmitted electron wave
$\psi=Ae^{ikx}$ at $\eps=\eps_{0}$ means the existence of a
solution of the Schr{\"o}dinger equation with $\psi\equiv0$ in one
lead. If the Hamiltonian is time-reversal symmetric, the complex
conjugated wave function should also be a solution with the same
boundary condition and $\eps=\eps_{0}^*$, hence $Im\eps_0\equiv
0$. The phase changes abruptly by $\pi$ when the real energy
$\eps_0$ crosses the Fermi energy.  [For the phase jumps
associated with the exact vanishing of the conductance there is no
way to distinguish between $+\pi$ and $-\pi$ jumps. This ambiguity
is resolved, after the phase jump acquires a some
width~\cite{SIKondo,Silva02}, due to e.g. a finite temperature.]

\begin{figure}[t]  \label{sequence}
\includegraphics[width=1.05\textwidth]{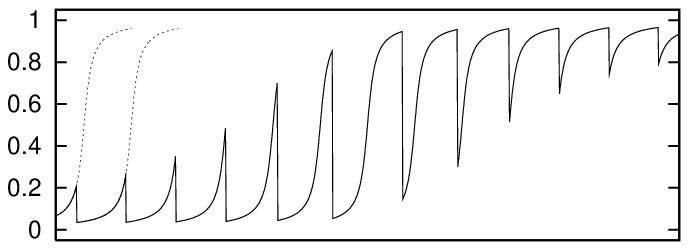}

\vspace{-1.25cm}

\includegraphics[width=1.05\textwidth]{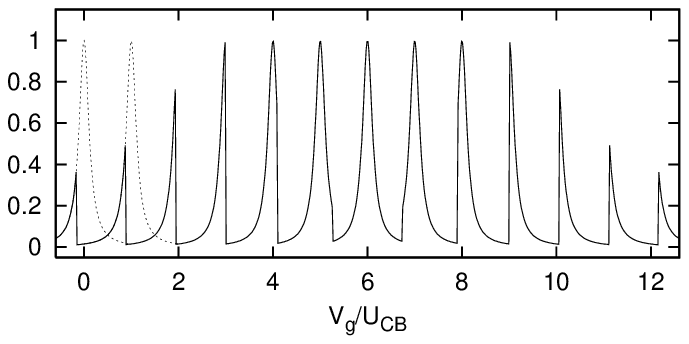}
\caption{A sequence of charging resonances described (exactly for
spinless electrons) by the theory of Ref.~\cite{SIPRL}. Upper
frame: the phase in units of $\pi$; lower frame: the transmission,
or conductance in units of $e^2/h$. Note that both the number of
repeated resonances, where the occupation switching takes place in
the CB valley, and the number of asymmetric peaks, where the
narrow level is charged inside the broad resonance, are
parametrically large, $\sim \Gamma/\Delta$. [For this figure we
chose $U_{CB}/\Gamma=4$ and $U/\Delta=25$. Increasing the ratio
$\Gamma/\Delta$ would allow having a larger number of repeated
peaks in the central part of the figure.] The conductance was
calculated by cutting  the usual Breit-Wigner resonances (two of
them shown dashed) "with scissors", and shifting them. }
\end{figure}

For the occupation switch taking place not too close to the
charging resonances, detailed calculations of the conductance and
phase behavior, may be done within the so called "elastic
cotunnelling" approximation, as we shall describe in Sec.~4.2.
Resulting from the destructive interference of quantum amplitudes,
the exact vanishing of the transmission amplitude here has some
similarity to the celebrated Fano zero. The broadening of the
phase lapses in the case of broken time reversal symmetry was
investigated in Ref.~\cite{Lee06}.

The level-occupation switching guarantees  the universal
occurrence of the phase jump. Otherwise, it depends on the
relative signs of the matrix elements connecting the states of the
dot to the leads~\cite{Silva02,Hersh,Oreg1}. For example, the
zeroes of the amplitude do not occur in the one dimensional case,
but should be found in about a half of the spin zero valleys in a
two-dimensional QD. While interesting effects can be found by
adjusting these signs, we believe that a more universal mechanism
is called for.

When there are a number of narrow levels in the vicinity of the
broad level, the above process can repeat many times. From
Eq.~(\ref{8}), we see that the number of narrow levels which switch
occupations with the broad one is on the order of ($\sim
(\Gamma/\Delta) ln(U_{CB}/\Gamma)$). Thus the above number of
consecutive resonances are due to the transition via one and the
same level $N$. Obviously, the conductance and the phase shift
(including the $\pi$ jumps) will be very similar for all these
resonances, as long as they occur near the middle of the CB valley.
When the occupation switch occurs close enough to a CB peak, the
shape of that peak will also be deformed (see below). For other
peaks, the phase will increase in a Breit-Wigner way by $\pi$ around
each resonance and sharply jump down by $\pi$ at the switching point
between the resonances. There is really no need for "further
calculation" of this phase behavior.

An example of such a series of consecutive resonances is shown in
Fig.~2. We note the typical increasing asymmetry of the
conductance peaks while going away from the center of the CB
valley. A peak asymmetry with a qualitatively similar structure
has in fact been observed by Lindemann~et.~al.~\cite{Lindemann}.
It was interpreted as due to the conductance being determined by a
special "bouncing ball" state having a width much larger than the
others~\cite{Hacken}. The mechanism invoked appears however to be
different from the one described here.

For the tails of series of peaks shown in Fig.~2, the occupation
switch
takes place within the charging resonance. With increasing
relative peak number, $|j-N|\gg \Gamma/\Delta$, the usual narrow
Breit-Wigner peaks in the conductance due to a population of level
$j$ should be formed here.
The shapes of these peaks, as well as (possible) Fano-type
conductance zeroes and $\pi$-phase drops will be strongly affected
by the many-body orthogonality catastrophe and their investigation
goes far beyond the scope of this paper.

For electrons with spin, the Breit-Wigner-related
formula~(\ref{4}) does not work, but the perturbative formulas
such as~(\ref{E0}) and~(\ref{5}) may be still used away from the
resonances. We will discuss the new effects caused by spin in
section~\ref{spin}. Before that we explain in the next subsection
why our model is likely to apply for a generic (smooth) QD with
mixed classical dynamics.

\begin{figure}
    \centering
\includegraphics[width=1.\textwidth,clip]{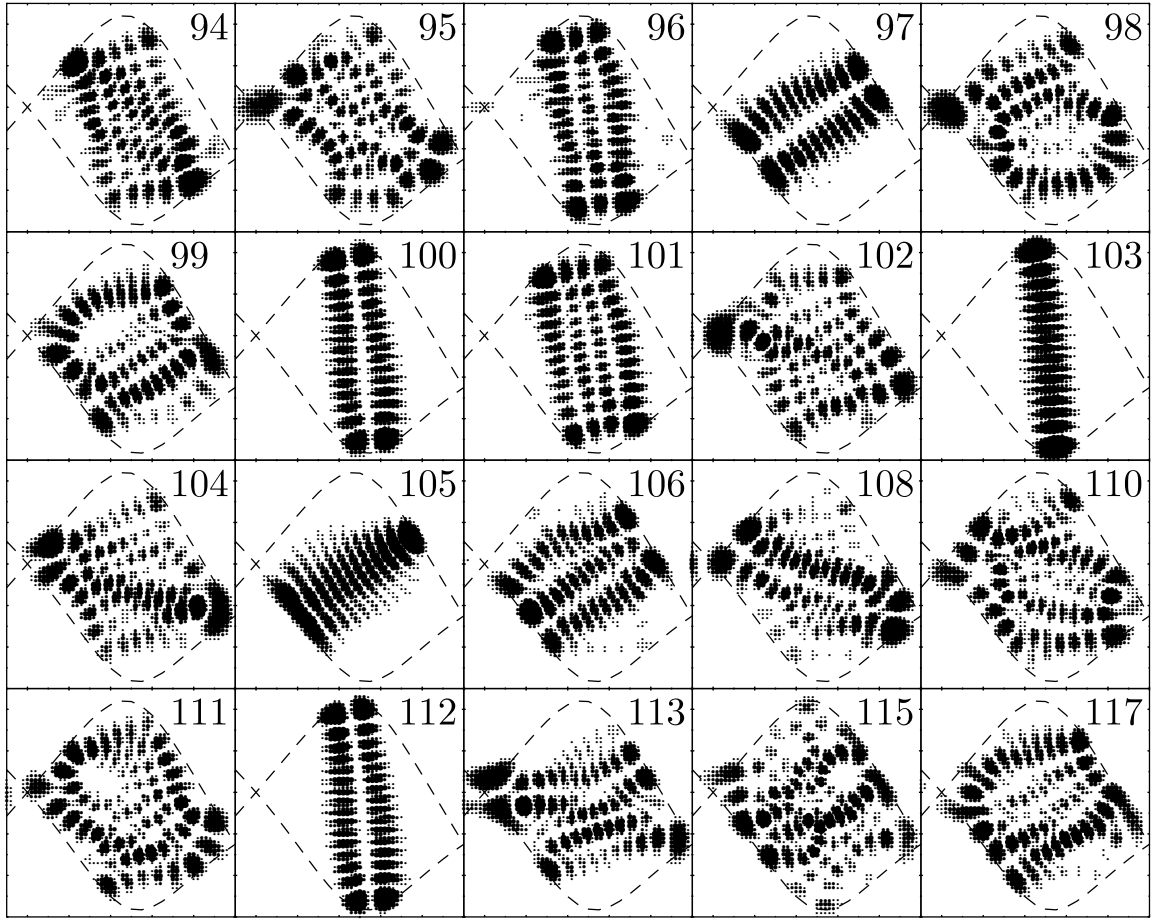}
\caption{The density of electrons in the dot at the resonances
coupled to a single-channel wire(attached from the left). The line
$V=0$ is shown dashed. Numbers correspond to the number of the
level in the QD. About $95\%$ of the norm of the wave function in
the dot is shown. The ``twin copies'' (such as 100 and 112) of
levels 94, 96, 101, 103 are not shown.}
\end{figure}

\subsection{Justification of the model}

We now discuss the conditions for obtaining levels with strongly
varying widths in the~QD.

An example of a system for which the widths differ drastically is
the integrable QD~\cite{Hackenbroich,Gefen}. However, it is hard
to believe that the large ($N_e\sim 100 \div 1000$) QD may be even
close to integrable. Nevertheless, at least in classical
mechanics, a considerable gap is left between integrable and fully
chaotic systems. Even in a nonintegrable dot two kinds of
trajectories - quasi-periodic and chaotic - may coexist. In this
case, in 2-dimensions any trajectory (even a chaotic one!) does
not cover all the phase space  allowed by energy conservation.
Consequently, the corresponding wave functions do not cover all
the area of the QD. If such a regime is realized in QD-s, it
easily explains why the widths of the resonances may vary by
orders of magnitude. Moreover, many other features of such a QD
may differ strongly from those of the chaotic QD~\cite{Stopa}. An
explicit numerical example, which supports the existence of such a
regime is given below~\cite{SIPRL}. We require the QD not to be
fully chaotic (neither do we require an integrable QD). It is not
clear, how chaotic was the dot used in the experiment. However,
the QD containing $\sim 200$ electrons was $\sim 50$ times smaller
than the nominal elastic mean free path. Thus, disorder should not
have been essential for the dynamics of the electrons.

To {\it illustrate} the relevance of the model Eq.~(\ref{Ham})
with single strongly coupled level we  performed numerical
simulations for a model QD of a size $l$ with a simple polynomial
potential (a smooth QD coupled to two leads)
 \bq\label{v}
V(x,y)= - 4x^2\left(1-\fr{x}{l}\right)^2 + \left(y+
\fr{x^2}{4l}\right)^2 \left(1+8 \left(\fr{x}{l}-\fr{1}{2}\right)^2
\right) .
 \ee
Due to the strong mixing of the $x$ and $y$ coordinates the dot is
expected to be nonintegrable, but, similarly to the experimental
geometry~\cite{Heiblum2}, it is approximately symmetric. For
simulations we considered the QD on the lattice and used $l=10$
which was equivalent to $50$ lattice spacings. The kinetic term is
given by the standard nearest neighbor hopping. Below we present
the results of calculations with the hopping matrix element
$\tau=18$ which corresponds to the dot with $\sim100$ electrons or
$\sim200$ if the spin is included (similar numbers to those in the
experiment). We have used the potential $V$ of Eq.(\ref{v}) for
$0<x<l$. The lead formed by the potential $V=3y^2$ was attached at
$x<0$ and a hard wall was put at $x=l$. Within the energy interval
$1.5<\eps<4.7$ exactly one mode may propagate along the lead. The
analysis of solutions of the Schr\"{o}dinger equation within this
interval allowed us to find the positions and widths of
quasi-stationary levels in the dot. As we expected, the widths
fluctuate very strongly from level to level (by many orders of
magnitude). In particular the widths of two levels $\# 102$ and
$\# 108$ exceed sufficiently the level spacing $\Gamma/\Delta\sim
6$ (the number of states doubled due to spin). The origin of the
hierarchy of widths becomes clear from fig.~1, where we have
plotted $|\psi|^2$ in the QD for (real) $\eps$ at the top of
corresponding resonances. The quantized version of different
variants of the classical motion may be found in this figure. The
narrowest level $\# 103$ corresponds to a short stable transverse
periodic orbit. Other broader levels, such as $\# 96,106$, may be
considered as the projections of the invariant tori corresponding
to quasi--periodic classical motion. This classical trajectory
reaches the line $V(x,y)=\eps$ only at few points. The candidates
for chaotic classical motion (e.g. $\# 110$) also correspond to
relatively broad resonances. Even in this case only a part of the
QD is covered by the trajectory. For the most coupled levels $\#
102$ and $\# 108$ the area covered by the trajectory touches the
lead by its corner.

Moreover, two well-coupled trajectories contribute to the level $\#
102$. This is seen from the fig.~2 where we show also the $|\psi|^2$
at the left and right wings of this resonance. One contribution
corresponds to the strongly coupled quasi-periodic trajectory
(left), having the ``turning point'' $V(x,y)=\eps$ just at the left
contact. The other contribution comes from the true periodic
trajectory (right). Two quantum states in the dot become mixed via
interaction with the wire and form one broad ($\# 102$) and one
almost decoupled ($\# 104$) resonance~\cite{Zelevinsky}. [This
mechanism was also found and investigated in ref. \cite{vO} and the
relationship to the Dicke effect \cite{Dicke} highlighted. It will
be further discussed at the end of this subsection.] We have
repeated the calculations several times for slightly different $V$
and in a broad range of variation of the hopping. Typically, we saw
resonances of very different widths and the origin of the broadest
peaks was explained by simple classical arguments. The explicit
example also allows to clarify the range of validity of our method.
In addition to the evident conditions $\Delta\ll\Gamma_N\ll U_{CB}$,
we should also have $\Gamma_N\ll\Delta\sqrt{N_e}$, $N_e$ being the
total number of electrons in the QD. For
$\Gamma_N\sim\Delta\sqrt{N_e}$ even the few broad levels start to
overlap and the CB is expected to be eliminated.

\begin{figure}
    \centering
\includegraphics[width=0.8\textwidth,clip]{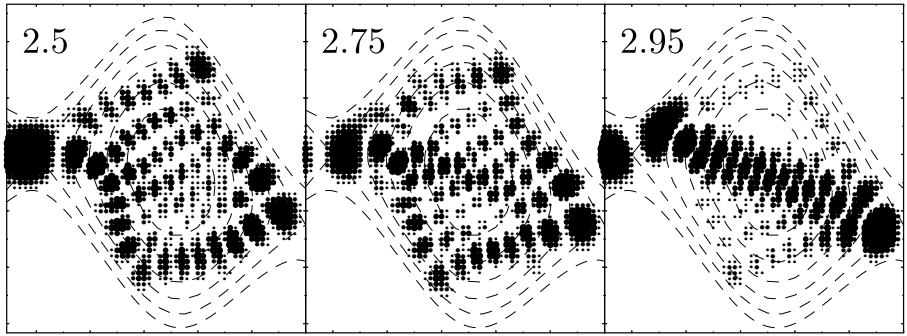}
\caption{Decomposition of the level $102$ into parts corresponding
to simple classical
trajectories. Numbers are the energies for which the figures
were done.}
\end{figure}

Taking into account the different sensitivities of longitudinal and
transverse modes to the plunger~\cite{Hackenbroich,Gefen} may allow
to keep our broad level $\eps_N$ even longer within the relevant
strip of energy.  This may provide an explanation for the even
longer sequences of resonances accompanied by the $-\pi$ jumps.

In a more refined approach, adding new electrons into the QD
should cause a change of the self-consistent potential $V(x,y)$.
The total energy of the dot and the wire will be lowered in the
presence of strongly coupled levels.  This may cause the potential
of the QD to automatically adjust to allow such levels, which will
support our explanation of the experiment of Ref.~\cite{Heiblum2}.

The Dicke-type mechanism for the dynamic generation of anomalously
broad levels was used for many years  in nuclear
physics~\cite{Zelevinsky}. In QDs the attention to this mechanism
was recently attracted by Ref.~\cite{vO}. The latter work also
generalized the treatment to several channels. When several levels
are coupled to a continuum and their widths are larger than their
spacings, the couplings among them mediated by the continuum
become important (the same coupling was alluded to as causing the
avoided crossing of levels $1$ and $0$ in section 2.1). One now
has to diagonalize the (nonhermitean, due to the finite lifetime
of the levels) effective Hamiltonian matrix. In the simplest case,
when all $n\gg 1$ initial levels are degenerate, one of the
resulting levels (the "superradiant" one \cite{Dicke}) will
acquire an $O(n)$ width, while the $n-1$ remaining levels will
become very narrow. This will still hold qualitatively when the
level spacings  do not vanish, but are smaller than the widths.
Such a generation of one broad level was also found in a more
general treatment, including the interactions and employing
renormalization group~(RG) approaches, in
Refs.~\cite{KarrFano,KarrNJP}.

We notice however, that our numerical example~\cite{SIPRL},
described above, showed that it may not be easy to employ the
Dicke mechanism in its full strength  in a QD described by a
realistic 2-dimensional potential. We were able to observe a
formation of the superradiant state from two close resonances
(levels $\#102$ and $\#104$ in Figs.~3 and 5). However, to arrange
the situation with several overlapping resonances, necessary for
this mechanism, is difficult since the averaged level width in
realistic QD appears to be always smaller than the level spacing.

\begin{figure}[t]
    \centering
\includegraphics[width=1.\textwidth,clip]{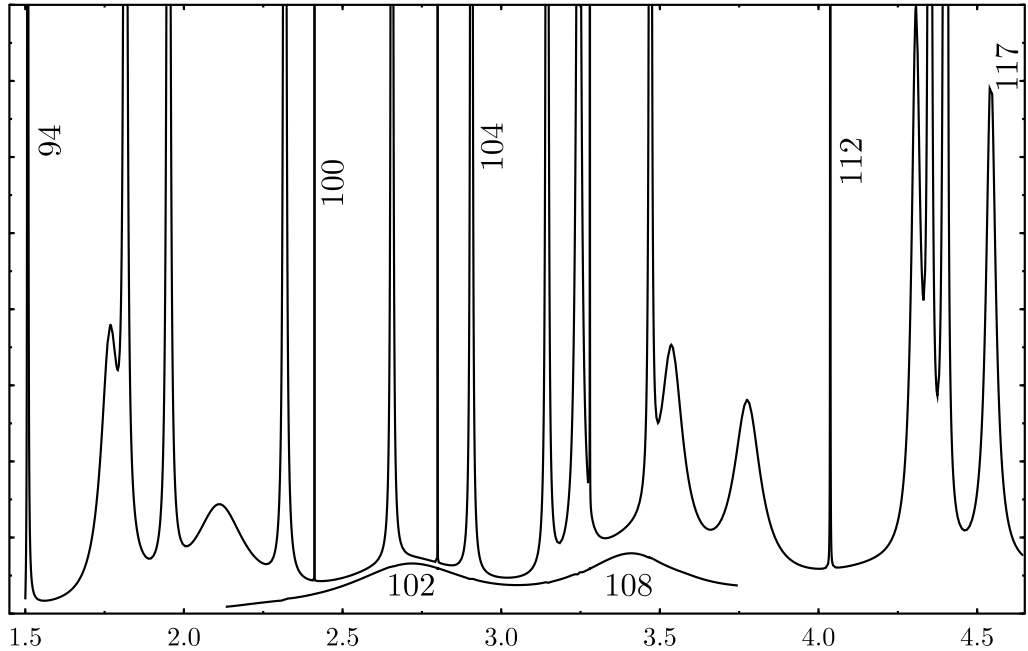}
\caption{The energy dependence of the probability to find the
electron in the QD for fixed incoming current. Numbers of some of
the levels in the dot are shown explicitly. The lower curve is the
result of subtraction of the narrow resonances, which allowed to
resolve the strongly coupled levels $\# 102$ and $\# 108$.}
\end{figure}

\section{Effects of the electron spin} \label{spin}

The goal of this section is to consider the new features which
appear if one goes beyond the simplest model of the occupation
switching considered in Sec.~2. This section  will be concerned with
the effect of the electron spin. Since adding the spin leads to a
significantly more complicated picture,  we consider in subsections
3.1 and 3.2 the case of a QD with only two levels, one well coupled
and one negligibly narrow. [These two subsections present the
results found in Ref.~\cite{SIPRB2002}.] The experimental tendency
towards producing small few-electron QD-s also encourages one to
investigate systems with small number of levels. The case of many
levels with spin will be discussed in Sec.~3.3.

\subsection{Two levels with spin, level-occupations and the ground
state energy}

Charging effects with "spinful" electrons are described by the same
tunnelling Hamiltonian Eq.~(\ref{Ham}), where now all summations
include the summation over the spin index as well. We consider a QD
with two levels, whose energies depends on a gate voltage $V_g$ and
level splitting $\Delta$ as in Eq.~(\ref{1})
 \bq\label{gate}
\eps_1=-V_g \ \ , \ \ \eps_2 =-V_g +\Delta  .
 \ee
Only one level, $\eps_1$, is well coupled to the leads, having the
width (\ref{2})
 \bq\label{Gamma}
\Gamma_1\equiv \Gamma
> \Delta \ \ , \ \  \Gamma\ll U_{CB}.
 \ee
The dot is charged by one electron at $V_g\approx
0,U_{CB},2U_{CB}$ and $3U_{CB}$.

Our first aim will be to find the ground state of the
system~(\ref{gate},\ref{Gamma}) at different values of $V_g$ in
the limit $\Gamma_2\rightarrow 0$. Similarly to Sec.~2, let us
denote by $E^{(0)}_{tot}, E^{(1)}_{tot}, E^{(2)}_{tot}$ the total
energy of the lowest state of the QD interacting with the leads,
with the narrow level populated by, respectively, $0,1$ and $2$
electrons (more precisely, $E^{(i)}_{tot}$ is defined as the
eigenenergy of the Hamiltonian~(\ref{Ham}) minus the trivial
energy of the electrons in the leads $\sum\eps(k)\langle a^+_k
a_k\rangle $). The functions $E^{(i)}_{tot}(V_g)$ evolve smoothly
with the gate voltage and the (averaged) occupation number of the
broad level $1$ also changes continuously. For example, the branch
$E^{(0)}_{tot}$ corresponds to an empty level $1$ at $V_g<0$,
singly occupied at $0<V_g<U_{CB}$ and doubly occupied at
$U_{CB}<V_g$. For $t_2^{L,R}=0$ the functions $E^{(i)}_{tot}$
cross at some values of $V_g$, which lead to sharp changes of the
ground state.

With the use of perturbation theory in $t^{L,R}_1$ it is easy to
find $E^{(i)}_{tot}$ far from the charging peaks.  Below the first
resonance (at $V_g\ll-\Gamma$) the true ground state is evidently
$E^{(0)}_{tot}$. However, already here the virtual jumps of the
electrons from the wire to the level $1$ give rise to the
correction
 \begin{eqnarray}\label{3}
 &&E^{(0)}_{tot}=
2\sum_{i=L,R}\int_{-E_F}^0 \fr{|t^i_N|^2}{\eps-\eps_N}
\fr{dn^i}{d\eps} d\eps
= \fr{- \Gamma}{\pi} \ln\left( \fr{4E_F}{\eps_1} \right) ,\\
&& E^{(1)}_{tot}= \eps_2- \fr{\Gamma}{\pi} \ln\left(
\fr{4E_F}{\eps_1+U_{CB}} \right) ,\nonumber\\
&& E^{(2)}_{tot} = 2\eps_2 +U_{CB} - \fr{\Gamma}{\pi} \ln\left(
\fr{4E_F}{\eps_1+2U_{CB}} \right).\nonumber
 \end{eqnarray}
The extra factor $2$ in $E^{(0)}_{tot}$ compared to Eq.~(\ref{E0})
accounts for the spin. It is clear that $E^{(0)}_{tot}$ in this
region lies significantly below $E^{(1)}_{tot}$ and
$E^{(2)}_{tot}$. When the (increasing) voltage crosses the region
$|V|\sim\Gamma$, the dot is charged by the first electron.
However, this electron may stay in the dot on the level $1$
(described by $E^{(0)}_{tot}$) or on the level $2$
($E^{(1)}_{tot}$). Depending on what level is occupied the
perturbation theory gives, in this range of $V_g$
 \begin{eqnarray}\label{33}
&&E^{(0)}_{tot}= \eps_1 -\fr{\Gamma}{2\pi} \left\{ \ln\left(
\fr{4E_F}{|\eps_1|} \right) +\ln\left(
\fr{4E_F}{\eps_1+U_{CB}} \right) \right\} ,\\
&&E^{(1)}_{tot}= \eps_2 -\fr{\Gamma}{\pi} \ln\left(
\fr{4E_F}{\eps_1+U_{CB}} \right) , \nonumber\\
&&E^{(2)}_{tot}= 2\eps_2 +U_{CB} - \fr{\Gamma}{\pi} \ln\left(
\fr{4E_F}{\eps_1+2U_{CB}} \right).\nonumber
 \end{eqnarray}
The first logarithm in $E^{(0)}_{tot}$ accounts for the virtual
jumps of an electron from the level $1$ in the dot to the wire.
The other logarithms correspond to  virtually adding the second
electron to the dot (having $\eps_1+U_{CB}$ or $\eps_1+2U_{CB}$
instead of $\eps_1$ in the denominator). The two levels
$E^{(0)}_{tot}$ and $E^{(1)}_{tot}$ cross at a gate voltage
(compare to Eq.~(\ref{8}))
 \bq\label{88}
V_g=V^{I}=\fr{U_{CB}}{\exp\{ -2\pi\Delta/\Gamma\} +1} \ .
 \ee
This result is valid for both signs of $\Delta$. For $\Gamma
\gg|\Delta|$, Eq.~(\ref{88}) reduces to $\eps_1\approx-U_{CB}/2$.
At $V_g=V^{I}$ the electron in the dot jumps from the broad level
to the narrow one.

In the second valley, $U_{CB}<V<2U_{CB}$ the dot is charged
already by two electrons, which again may populate  the two doubly
degenerate levels in the dot in different ways. Here one finds,
 \begin{eqnarray}\label{33333}
E^{(0)}_{tot}&=& 2\eps_1+U_{CB} -\fr{\Gamma}{\pi}  \ln\left(
\fr{4E_F}{|\eps_1+U_{CB}|} \right) \ , \\
E^{(1)}_{tot}&=& \eps_1+\eps_2+U_{CB} -\fr{\Gamma}{2\pi}
\left\{\ln\left( \fr{4E_F}{|\eps_1+U_{CB}|}\right) +\ln\left(
\fr{4E_F}{\eps_1+2U_{CB}} \right)
\right\}\ , \nonumber\\
E^{(2)}_{tot}&=& 2\eps_2+U_{CB} -\fr{\Gamma}{\pi}  \ln\left(
\fr{4E_F}{\eps_1+2U_{CB}} \right) \ . \nonumber
 \end{eqnarray}
First of all, we see that just after the charging peak, at
$V_g>U_{CB}$, the true ground state is $E^{(0)}_{tot}$. This is in
contrast with the situation at $V_g<U_{CB}$, where the ground state
was $E^{(1)}_{tot}$~(\ref{33}). Thus we may conclude, that {\it
within the resonance} not only is  one electron gradually
transmitted from the wire to the level $1$ in the dot, but also {\it
a second electron} is taken from the narrow level $2$ to the broad
one $1$ (see fig.~6). In the limit $\Gamma_2\rightarrow 0$ this
second ``transfer'' is abrupt and takes place at some
$V_g=W^I\approx U_{CB}$ (for zero temperature, $T\equiv 0$). This
prediction of the possibility to have sharp features within the
charging resonance is {\it the main new effect caused by the spin}.
All the three energies (\ref{33333}) cross at the same value of gate
voltage (c.f. (\ref{88}))
 \bq\label{888}
V^{II}=U_{CB}+\fr{U_{CB}}{\exp\{ -2\pi\Delta/\Gamma\} +1}
 \ee
and the true ground state becomes $E^{(2)}_{tot}$. Now already two
electrons jump together from the broad level to the narrow one. At
$V_g>V^{II}$ (\ref{88}) the occupation of the quantum dot proceeds
in a fashion symmetric to the above. At the third peak the third
electron is added to the dot. Were the branch $E^{(2)}_{tot}$  the
stable one, this would have been the uncoupled electron at the
level $1$. However, $E^{(2)}_{tot}$ and $E^{(1)}_{tot}$ cross at
the top of the third peak (at $V_g=W^{II}=W^I+U_{CB}$) and the
ground state for the first half of last valley has a single
unpaired electron at the narrow level. Finally, at
 \bq\label{888dd} V^{III}=2U_{CB}+\fr{U_{CB}}{\exp\{
-2\pi\Delta/\Gamma\} +1}
 \ee
$E^{(2)}_{tot}$ and $E^{(1)}_{tot}$ cross and the broad level
becomes singly occupied again.  The fourth peak ($V_g\approx
3U_{CB}$) completes the charging of two levels in the quantum dot by
four electrons.

\begin{figure}
    \centering
\includegraphics[width=0.8\textwidth]{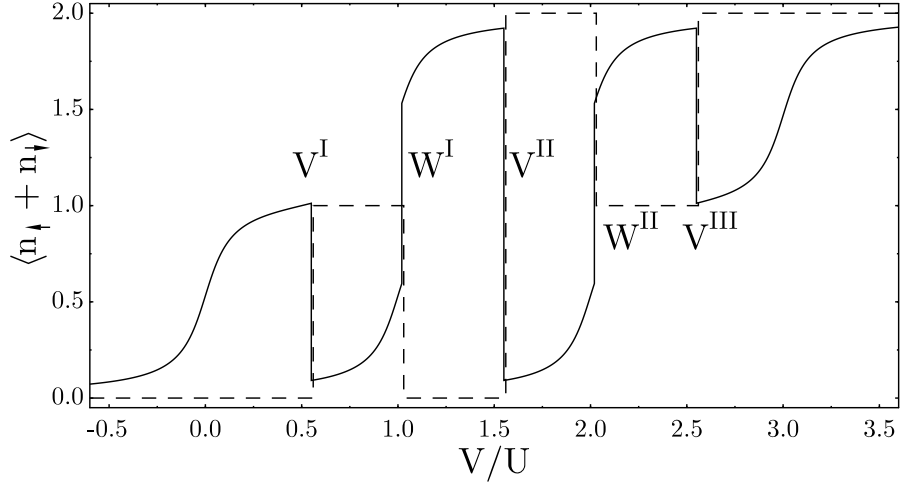}
\caption{Schematic drawing of the averaged occupation number
$\langle n_{\uparrow} + n_{\downarrow} \rangle$ for the broad
level $1$ (solid) and narrow level $2$ (dashed). The gate voltage
$V$ is measured in units of the charging energy $U_{CB}$. The
charging resonances are at $V_g/U_{CB}\approx
0,1,2,3$.}\label{FigOccupation}
\end{figure}

To summarize the discussion of this subsection, we show
schematically in Fig.~\ref{FigOccupation} the averaged occupation
numbers of our two levels $\langle n_{1\uparrow}+
n_{1\downarrow}\rangle$ and $\langle n_{2\uparrow}+
n_{2\downarrow}\rangle$. The four charging resonances correspond to
$V_g/U_{CB}\approx 0,1,2,3$. One may see on the figure many abrupt
changes in the occupation of both broad and narrow levels. The
occupation number of a given level is not measured directly in the
experiment. However, the appearance of an occupation switching may
be seen in the signal of a suitable QPC detector \cite{QPC} as we
describe in Sec.~\ref{subsectionQPC}. In the following subsections
we discuss the transport properties of the QD with a single broad
level for "spinful" electrons.

\subsection{Two levels with spin. Conductance}\label{secConducTwoSpin}

\begin{figure}[t]
    \centering
\includegraphics[width=0.8\textwidth]{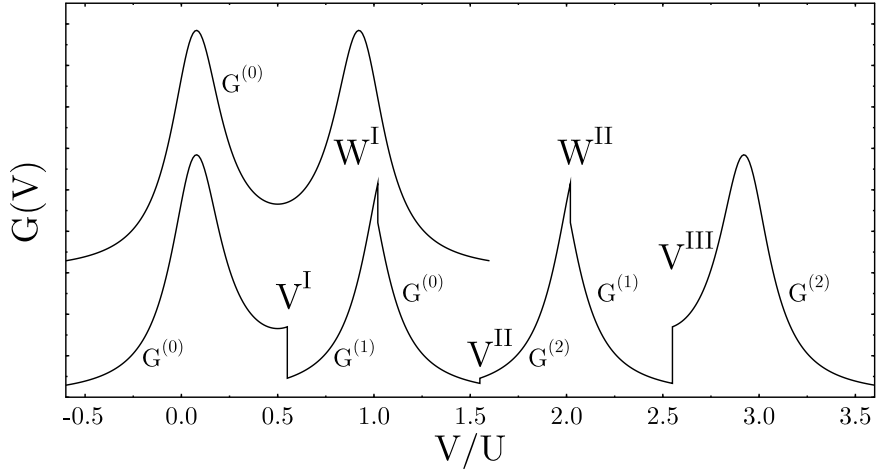}
\caption{The gate voltage dependence (schematic) of the conductance
$G$. The upper curve shows $G^{(0)}$ in the case of only one (broad)
level in the quantum dot (vertically offset for clarity). The lower
curve depicts the conductance given by the different $G^{(n)}$ in
various regimes. Sharp features at the two peaks and three valleys
are seen.}\label{FigCondSpin}
\end{figure}

The zero bias conductance $G$ of our two-level quantum dot is shown
schematically in  Fig.~\ref{FigCondSpin}. In the limit of an
"invisible" level $2$ ($\Gamma_2 \rightarrow 0$), we may introduce
three conductances $G^{(0,1,2)}$ corresponding to empty,
singly-occupied and doubly-occupied narrow level (characterizing the
three "ground state" energies $E^{(0,1,2)}_{tot}$ of the previous
section). The role of the electrons in the level $2$ reduces in this
case to simply raising  the current-transmitting level $1$ via the
Coulomb repulsion. Thus
 \bq\label{rising}
G^{(0)}(V_g)=G^{(1)}(V_g+U_{CB})=G^{(2)}(V_g+2U_{CB}) \ .
 \ee
We have shown schematically the function $G^{(0)}$ on the same
Fig.~\ref{FigCondSpin} (vertically offset). In the limit
$\Gamma_2\ll\Gamma_1$ the curve for a two-level dot is obtained
simply by cutting and horizontally shifting the parts of the curve
for the single-level dot, in agreement with Eq.~(\ref{rising}), as
shown in the figure. Thus the relations Eq.~(\ref{rising}) allow
one to describe the singular behavior of the conductance even
without an explicit calculation of $G^{(0)}$.

The sharp features seen in Fig.~\ref{FigCondSpin} will be smeared at
finite temperature (see examples in Ref.~\cite{SIPRB2002}). To
resolve them one needs to measure the conductance at $T\ll\Gamma$.
If the temperature is still large compared to the Kondo temperature,
$T_K=(U_{CB}\Gamma/2)^{1/2}e^{\pi \eps(\eps+U_{CB})/2\Gamma
U_{CB}}$, the conductance away from the resonances may be found by
calculating the transmission tunneling amplitude in the second order
of perturbation theory ("elastic cotunnelling"). This gives,
 \begin{eqnarray}\label{G1}
G^{(0)}=2\fr{e^2}{h} \fr{\Gamma_L\Gamma_R}{\eps_1^2}, \ \ \ \ \ \
\ \ \ {\rm at} \ \
V_g<0,\\
G^{(0)}=\fr{e^2}{h}\Gamma_L\Gamma_R \left[
 \left(\fr{1}{\eps_1}-\fr{1}{\eps_1+U_{CB}}\right)^2
 +\fr{1}{\eps_1^2}+\fr{1}{(\eps_1+U_{CB})^2}
\right], \nonumber\\
\ \ \ \ \ \ \ \ \ \ \ \ \ \ \ \ \ \ \ \ \ \ \ \ \ \ \ \ \ \ \ \ \
{\rm at} \ \
0<V_g<U_{CB},\nonumber \\
G^{(0)}=2\fr{e^2}{h} \fr{\Gamma_L\Gamma_R}{(\eps_1+U_{CB})^2}, \ \
{\rm at} \ \ U_{CB}<V_g. \nonumber
 \end{eqnarray}
The first line in Eq.~(\ref{G1}) is for an empty dot. The second
line describes the current through the QD having one electron on
the level $\eps_1$. The current in this case is a sum of three
contributions~\cite{SIPRB2002} accounting respectively for the
spin-flip processes, transmission of the electron with the spin
parallel to the spin of the electron in the QD and transmission of
the electron with antiparallel spin. Note that the spin-flip
contribution vanishes at $U_{CB} = 0$, as it should. The last line
in Eq.~(\ref{G1}) is for the case of two electrons on the level
$\eps_1$.

Two kinds of sharp features are seen in Fig.~\ref{FigCondSpin}.
First, both a cusp and a jump  take place at the peaks: $W^{I}$,
where the curve $G^{(1)}$ is replaced by $G^{(0)}$ and $W^{II}$,
where $G^{(2)}$ is replaced by $G^{(1)}$. The jump vanishes for
$\Delta\ll\Gamma$, but the pronounced cusp survives even for
$\Delta=0$. Currently, no theory  exists which would describe the
fate of these singularities at $T\rightarrow 0$ and finite (small)
coupling to the leads of the level $\eps_2$.

Besides the above, there are three jumps of $G(V_g)$ in the
valleys. The values of the conductance at the borders of the first
jump, at $V_g=V^I\pm 0$ are (assuming $\Delta \ll\Gamma$)
 \bq\label{jump} G\approx 48\fr{e^2}{h}
\fr{\Gamma_L\Gamma_R}{U_{CB}^2} \ \ {\rm and} \ \ G\approx
16\fr{e^2}{h} \fr{\Gamma_L\Gamma_R}{U_{CB}^2} \ .
 \ee
The contribution from the spin flip processes for $V_g<V^I$ is
twice that from the elastic processes. Therefore, the conductance
drops near $V_g=V^I$ by a factor of 3. At $V_g=V^{II}$ the two
electrons jump from the broad to narrow level. The discontinuity
at $V^{II}$, which follows from the small difference in the
probability of the electron-like and hole-like processes, vanishes
for $\Delta\ll\Gamma$ (restoration of particle-hole symmetry at
$\Delta=0$). Examples which find the fine structure of such jumps
at $\Gamma_2\neq 0$ will be given below.

\subsection{Many levels with spin}\label{secManylevelsSpin}

\begin{figure}[t]
    \centering
\includegraphics[width=1.05\textwidth]{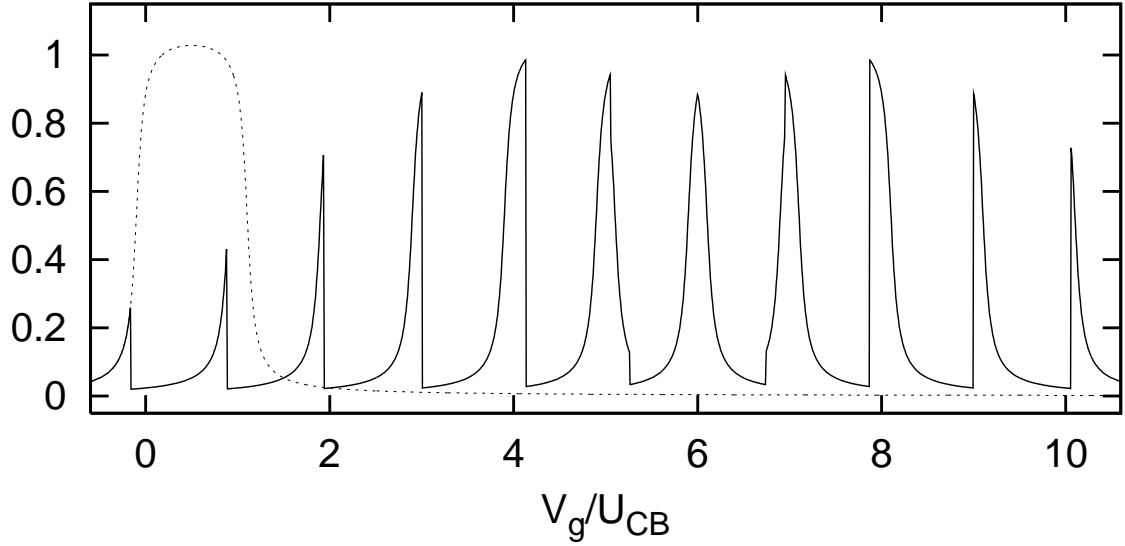}
\caption{A series of  conductance resonances in a QD with a single
broad level with spin. Here in the first five peaks electrons first
try to populate the broad level with either spin up or spin down and
then (before the charge at the broad level reaches $1\times e$) are
thrown away after the charging of the narrow level. Similarly the
last three peaks (we don't show more because the picture is
symmetric) the lead-electron jumps onto the narrow level in the QD,
which raises the energy of occupied broad level and increases the
transport through it. In the three peaks in the middle, one electron
jumps from the narrow to the broad level in the QD inside the
charging resonance. The jumps in the tails of the peaks here
correspond to depopulation of both spin states of the broad level.
The thin dashed line shows the (schematic) conductance for the
Anderson impurity model at zero temperature, including the increase
of the conductance upon adding the first electron, the Kondo plateau
and the conductance drop after adding the second electron. The
conductance for the multilevel QD is obtained by gluing and
repeating the pieces of this curve.}\label{FigManylSpin}
\end{figure}

A straightforward generalization of the theory presented in previous
two subsections allows one to find the conductance (and level
occupancies) for a multilevel QD with a single broad level,
$\Gamma\gg\Delta$, for "spinful" electrons. The resulting
conductance behavior is shown in Fig.~\ref{FigManylSpin} (compare
with Fig.~2). Similarly to the case of 2-level dot, considered in
Sec.~\ref{secConducTwoSpin}, the conductance curve may be found by
cutting and gluing the parts of the curve $G^{(0)}(V_g)$ describing
a QD with single (spinful) level. In contract to
Fig.~\ref{FigCondSpin} we now consider the case of zero temperature
($T\ll T_K$). Thus the "generating function" for conductance of
multilevel QD is now the conductance at $T=0$ for the Anderson
impurity model~\cite{Anderson}, shown by thin dashed line in the
figure. [An exact Bethe ansatz solution for this model~\cite{Bethe}
gives $G^{(0)}(V_g)$ explicitly.]

The series of peaks shown in Fig.~\ref{FigManylSpin} contains three
large, $\sim\Gamma/\Delta$, sets with qualitatively different
charging scenarios. First comes a sequence of peaks
($V_g/U_{CB}\approx 0,1,2,3,4$ in the figure) having one occupation
switch per charging event. At the left wing of each resonance here
the broad level in the QD is empty and is gradually populated with
the increasing $V_g$. Somewhere inside the charging resonance it
becomes energetically favorable to populate one of the narrow levels
in the dot by single electron. The broad level is depopulated at
this moment and its energy is raised, $\eps_N\rightarrow \eps_N
+U_{CB}$. A reliable description of the fine structure of such an
occupation switching for a finite coupling of the narrow levels to
the leads, remains a challenging theoretical problem (similarly to
the case of spinless electrons of Sec.~2.1).

With further increase of the gate voltage the occupation
switching, for a series of $\sim\Gamma\ln(U_{CB}/\Gamma)$ peaks,
splits into two parts. Now inside each peak one electron jumps
from the narrow to the broad level in the dot. Such events for
2-level case were considered in previous subsections ($V_g\approx
W^I$ and $V_g\approx W^{II}$). Thus within a single charging
resonance the well-coupled level in the dot becomes occupied by
two electrons. Between the resonances both electrons from the
broad level jump to narrow ones. This is where one will see the
$\pi$ jumps of the transmission phase in the CB valley.

Finally, for a series of $\sim\Gamma/\Delta$ peaks the charging
events start (left wing of the resonance) in the situation when the
broad level stays inside the QD ($\eps_N\approx -U_{CB}$) being
populated by two electrons. Then, at a certain voltage, an electron
from a lead jumps onto the QD narrow level. Repulsion from this
electron raises the broad level,
$\eps_N\rightarrow\eps_N+U_{CB}\approx\Gamma$, and partially
depopulates it. The occupation of the level $N$ is completed again
at the right wing of the resonance.

The important consequence of the scenario presented in this section
is the absence of the Kondo effect for a large series of charging
resonances, which might be easy to verify experimentally.

\section{Sharp features and charge detection}

In subsections 4.1 and 4.2 we considered the effects which appear if
one takes into account the finite width of the narrow level. Since
the particular features considered here do not depend on the spin
the discussion is narrowed to spinless electrons only.

\subsection{Sharp features: QPC detector signal}\label{subsectionQPC}

In this subsection we investigate the fate of the sharp features
in the CB valley considered before, if the narrow levels in the QD
acquire a small but finite widths. Let, as in Sec.~3.1, 1 be a
broad level and 2 be a narrow one in the QD. First, in a way
analogous to Eqs.~(\ref{2},\ref{Gamma}) we introduce the width of
the second level $\Gamma_2$ and the interlevel coupling
$\Gamma_{12} = 2\pi\sum_{i=L,R} t_1^i t_2^{i*} {dn^i}/{d\eps}$.
Consider a CB valley where there is one electron in the dot either
on the level 1, or on the level 2. The dynamics of QD is described
by the effective single-particle Hamiltonian, accounting for the
coupling to the leads, whose matrix elements are
($\eps_2-\eps_1=\Delta$)~\cite{SIPRB2002}
 \begin{eqnarray}\label{HamEff}
H_{11}&=&\eps_1-\fr{\Gamma_1}{2\pi}\ln\left(
\fr{\eps+U_{CB}}{|\eps|}\right) \, , \,
H_{22}=\eps_2-\fr{\Gamma_2}{2\pi}\ln\left(
\fr{\eps+U_{CB}}{|\eps|}\right)
 , \nonumber\\
H_{12}&=&-\fr{\Gamma_{12}}{2\pi}\ln\left(
\fr{\eps+U_{CB}}{|\eps|}\right) = H_{21}^* \  .
 \end{eqnarray}
Here $H_{11}$ and $H_{22}$ are nothing more than the renormalized
single particle energies $\eps_1$ and $\eps_2$, similar to
Eqs.~(\ref{5},\ref{6}) minus a proper constant. In the argument of
the logarithm $\eps$ may either be $\eps_1$ or $\eps_2$, since
$\eps+U_{CB},|\eps|\gg \Delta$. For simplicity we consider the
time reversal symmetric case, $H_{21}^*=H_{21}$.
Eq.~(\ref{HamEff}) is valid for an arbitrary ratio of couplings of
two levels to the leads. In the general case however, the width of
the level switch may become of order $U_{CB}$ (see
Eq.~(\ref{deltaGdel}) and remark below it). Below we consider the
sharp switch case, $\Gamma_1\gg\Gamma_2$,
$\Gamma_{12}\sim\sqrt{\Gamma_1\Gamma_2}$.

\begin{figure}
    \centering
\includegraphics[width=0.8\textwidth,clip]{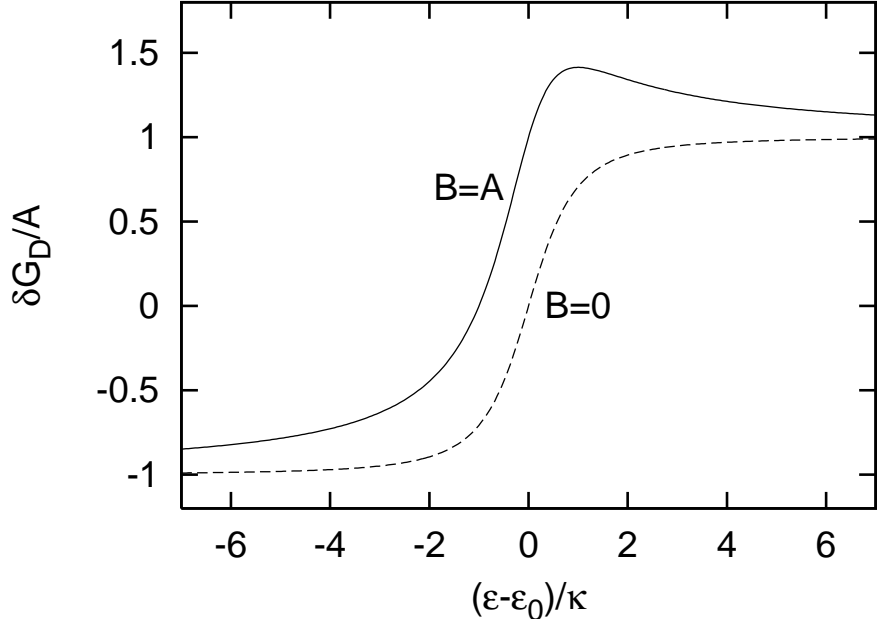}
\caption{The signal at the detector (QPC) for two values of the
parameters used in Eq.~(\ref{deltaGdel}). The dashed line shows the
case of vanishing cross-scattering amplitude ($B=0$). The solid line
shows the case then the scattering amplitude for the QPC electron
with flipping the state of the electron in the QD from 1 to 2,
equals to the difference of direct scattering amplitudes for the dot
electron in states 1 and 2 ($B=A$).} \label{FigQPC}
\end{figure}

For electrons with spin there are two species of the effective
Hamiltonian Eq.~(\ref{HamEff}) for each spin orientation. [If in
addition the QD has one electron for each spin, there are no
spin-flip transitions in the linear conductance.] In this and in the
next section we consider spinless electrons.

Examples of resolving the sharp features in the conductance based on
the avoided level crossing described by Eq.~(\ref{HamEff}) may be
found in Ref.~\cite{SIPRB2002}. Here we give only two examples,
which seemingly have not been covered in the existing literature.

Using a side quantum point contact (QPC) as a charge
detector~\cite{QPC} becomes now an efficient experimental tool for
finding the number of electrons in small QDs, complimentary to the
transport measurements. The same device may obviously be used to
detect the occupation switching~\cite{KG}.

The ground state of the system described by the
Hamiltonian~(\ref{HamEff}) is a superposition of two states
 \bq\label{alphabeta}
|\tilde{1}\rangle=\alpha|1\rangle +\beta|2\rangle \ , \
\alpha,\beta=\pm\sqrt{\fr{1}{2} \mp
\fr{H_{11}-H_{22}}{2\sqrt{(H_{11}-H_{22})^2+4|H_{12}|^2}}}.
 \ee
For weak coupling between the QD and QPC the transmission
amplitude through the latter reads ($\delta \tau\ll \tau_0$)
 \bq\label{tQPC}
\tau=\tau_0+\delta t \ , \ \delta \tau=
(\alpha^2-\beta^2)\fr{\langle 1|R|1\rangle - \langle 2|R|2\rangle
}{2} +2\alpha\beta \langle 1|R|2\rangle,
 \ee
where $R$ is some single-particle operator and both the background
transmission amplitude $t_0$ and the matrix elements of $R$ weakly
depend on the gate voltage $V_g$. Thus we find the change in the
QPC conductance
 \bq\label{deltaG}
\delta G_D
=A\fr{H_{11}-H_{22}}{\sqrt{(H_{11}-H_{22})^2+4|H_{12}|^2}} +
B\fr{2H_{12}}{\sqrt{(H_{11}-H_{22})^2+4|H_{12}|^2}},
 \ee
where $A=\fr{e^2}{h}\tau_0(\langle 1|R|1\rangle - \langle
2|R|2\rangle)$ and $B=\fr{e^2}{h}2 \tau_0 \langle 1|R|2\rangle $. To
see explicitly the gate voltage dependence of $\delta G_D$ we
consider the case of a single broad level, $\Gamma_1\gg
\Gamma_{12}\gg \Gamma_2$. Now we may omit $\Gamma_2$ in
Eq.~(\ref{HamEff}) and use an expansion around the center of the
switch ($\eps_1=-V_g$)
 \bq
\eps_0=\fr{-U_{CB}}{\exp(-2\pi\Delta/\Gamma_1)+1},
 \ee
which allows us to write
 \bq
H_{11}-H_{22}\approx -\fr{\Gamma_1}{\pi U_{CB}}
\sinh\left(2\pi\fr{\Delta}{\Gamma_1} \right)(\eps_1-\eps_0) \ , \
H_{12}\approx\fr{\Gamma_{12}}{\Gamma_1}\Delta.
 \ee
Now, instead of Eq.~(\ref{deltaG}) one has
 \bq\label{deltaGdel}
\delta G_D
=\fr{(\eps_1-\eps_0)A}{\sqrt{(\eps_1-\eps_0)^2+\kappa^2}} +
\fr{\kappa B}{\sqrt{(\eps_1-\eps_0)^2+\kappa^2}} \ , \
\kappa=\fr{\Delta \Gamma_{12} U_{CB}}{\Gamma_1^2
\sinh\left({2\pi\fr{\Delta}{\Gamma_1}} \right)}
 \ee
The width of the occupation switch, $\kappa$, is always a fraction
of the charging energy, becoming $\kappa=\fr{ \Gamma_{12}
U_{CB}}{\Gamma_1}$ for $\Delta << \Gamma_1$. The function $\delta
G_D (\eps_1)$ is shown in Fig.~\ref{FigQPC}. Note the asymmetry of
the step in the QPC signal and the maximum on one side of the step
for $B\neq 0$.

\subsection{Sharp features: Phase drops}

While we declared in Sec.~2.1 that the phase lapses in the CB valley
for spinless electrons should be abrupt and equal $\pi$, the actual
phase drops in Fig.~\ref{sequence} are always somewhat smaller than
$\pi$. In this section we show how this exact $\pi$ is restored
after the coupling of the narrow levels is switched on (still we
assume $\Gamma_1\gg \Gamma_{12}\gg \Gamma_2$). We also continue
working with spinless electrons here.

\begin{figure}[t]  \label{Smallsequence}
    \centering
\includegraphics[width=.8\textwidth]{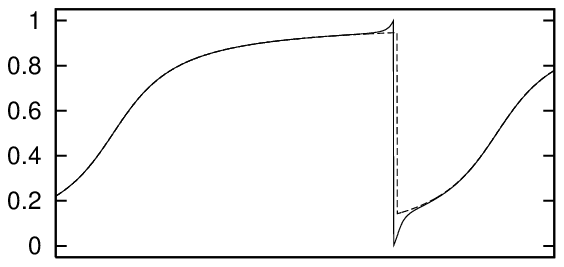}

\vspace{-1.15cm}

\includegraphics[width=.8\textwidth]{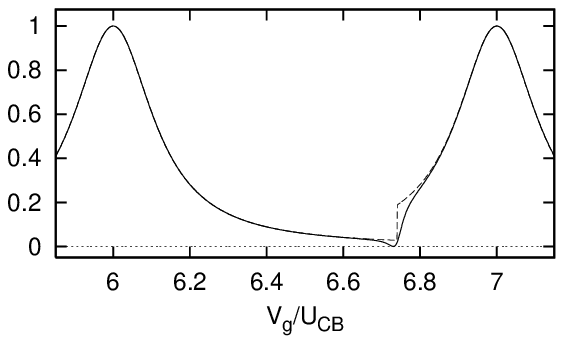}
\caption{The extended part of Fig.~2 showing the fine structure of
the conductance and the phase behavior at the occupation
switching, as described by Eqs.~(\ref{conductSwitch}) and
(\ref{phaseSwitch}). The phase jump now becomes exactly
$\Delta\phi=\pi$, while the conductance is smooth and vanishes at
a certain point. The value of $\kappa$~(\ref{deltaGdel}) is taken
to be $\kappa=U_{CB}/50$. The dashed lines show the bare curves
obtained "with scissors", as in Fig.~2 (with the abrupt jump in
the conductance and the phase jump $\Delta\phi<\pi$). }
\end{figure}

To find the conductance and phase behavior we first notice that
after having successfully employed the coupling of the narrow level
to smoothen the occupation switch, there is no need to consider the
contribution to the conductance from this small coupling. What we
have now is a QD with two levels $|\tilde{1}\rangle$ and
$|\tilde{2}\rangle$~(\ref{alphabeta}), each coupled to the leads via
tunneling matrix elements $\alpha t^{L,R}$ and $\beta t^{L,R}$. One
level, $|\tilde{1}\rangle$, is occupied by the electron (energy
$\eps<0$), the other level is raised due to electron repulsion
($\eps+U_{CB}$).

In the valley between two CB peaks, the conduction is dominated by
the  so-called elastic cotunneling (i.e.second order) processes
through the resonances at larger and smaller energies. For
example, tunneling first from the left lead to the dot (whose
relevant state must then be empty beforehand)  then from the dot
to the right lead. This "electron process"  is dominant near the
right-hand CB peak. The other possibility, dominant near the
left-hand CB peak, where the relevant dot level is full, involves
first a transition of the electron from the dot to the right lead,
then from the left lead to the dot. Therefore, the latter process
is termed a "hole process". For fermions, the hole process is
well-known to acquire an extra minus sign. In our case, where due
to the level-occupation switching, the two CB peaks are due to the
same level, the transmission amplitude {\em must} have opposite
signs in the two sides of the Coulomb valley. Thus the
transmission amplitude in the time-reversal symmetric case must
have a zero somewhere in-between. A simple calculation with the
use of Eq.~(\ref{alphabeta}) now gives (we assume $t^L=\pm t^R$).
 \bq\label{conductSwitch}
G=\fr{e^2}{8h}\left( \fr{(1-\xi)\Gamma_1}{\eps} +
\fr{(1+\xi)\Gamma_1}{\eps+U_{CB}}\right)^2 \ , \
\xi=\fr{\eps-\eps_0}{\sqrt{(\eps-\eps_0)^2+\kappa^2}}.
 \ee
To extract the transmission phase from this result we further notice
that since the couplings of two levels to the leads are proportional
we may perform a standard unitary rotation of the
leads~\cite{KondoT1},  $a_k\propto t^L a_k^L +t^R a_k^R$,
$a_k'\propto -t^R a_k^L +t^L a_k^R$, to define one coupled ($a_k$)
and one decoupled ($a_k'$) lead. All transport properties are found
from the $S$-matrix for the single coupled lead, $S\equiv
e^{2i\phi}$. In particular the transmission amplitude and
conductance are (again $|t^L|=|t^R|$)
 \bq\label{phaseSwitch}
\tau = \sin\phi e^{i\phi} \ , \ G=\fr{e^2}{h}\sin^2\phi.
 \ee
This allows us to relate the conductance~(\ref{conductSwitch}) and
the phase of the transmitted wave~$\phi$. Fig.~\ref{Smallsequence}
shows the conductance and the phase behavior found from
Eqs.~(\ref{conductSwitch}) and (\ref{phaseSwitch}).

An elastic cotunnelling treatment of transmission zeroes in
noninteracting QDs was presented in Ref.~\cite{Silva02}.
Ref.~\cite{Lee06} investigated the evolution of phase lapses in an
interacting QD including the case of broken time-reversal
symmetry.

\section{The sensitivity of the transmission phase to Kondo
correlations}

Among the observations~\cite{Kondo1,Kondo2,Kondo3,Kondo4,Kondo5}
of the Kondo effect~\cite{Hewson} in quantum
dots(QD)~\cite{KondoT1,KondoT2,KondoT3}, two
experiments~\cite{Yang1,Yang2} were devoted to the measurement of
the phase shift of the transmitted electron. These experiments
were aimed at a direct observation of the fundamental
prediction~\cite{Langreth,Nozieres} of the Kondo model: the $\phi=
\pi/2$ phase shift experienced by the scattered electron after the
spin of the impurity is screened into a singlet.

In agreement with the general predictions and the numerical
renormalization group calculations~\cite{Delft} for an Anderson
impurity~\cite{Anderson} in Aharonov-Bohm(AB) interferometer, the
development of a plateau of $\phi$ in the Kondo valley was seen in
Ref.~\cite{Yang1} (although the reported saturation value of the
phase was not $\pi/2$). A very important, unexpected feature of
the experimental results was the strong sensitivity of the phase
to Kondo correlations: the phase saturated at $\phi\approx const$,
while the conductance in the valley was still well below the
unitary limit, indicating the absence of Kondo screening.

Our  aim in this section will be  to explain this strong Kondo
effect in the phase \cite{SIPRL}. We will see that although the
nontrivial phase behavior is indeed governed by the Kondo physics,
the main changes of phase take place at temperatures
parametrically larger than $T_K$. In a sense, we show that the
phase changes in the regime where, although the spin of the
impurity is not screened, the running Kondo coupling constant
exceeds parametrically its bare value.

The transmission phase is measured, as before,  by embedding the
quantum dot into one arm of the two-wave AB
interferometer~\cite{Heiblum2}. Let $A_{s S}^{d}$ be the
transmission amplitude for an electron with spin projection $s$
through the QD having a spin projection $S$. Respectively, let
$A_{s}^r$ be the transmission amplitude through the second reference
arm. (Obviously, the spin- flip processes occurring in the QD and
not in the reference arm do not contribute to the interference.)
Now, the part of the current oscillating with the change of the
magnetic flux threading the interferometer takes the form
 \bq\label{AB}
G_{AB} \propto Re \sum_{s S}A_{s}^{r*}A_{s S}^{d} = Re
A^{r*}\sum_{s S}A_{s S}^{d} \ .
 \ee
We use the fact that (for the weak magnetic field used in the
experiment) the transmission through the reference arm does not
depend on spin. By measuring the relative phase of the AB
oscillations at different values of the gate voltage one extracts
the information about the transmission phase. We see that in fact
in the interference experiment not a single amplitude is measured,
but the sum of all possible amplitudes corresponding to different
orientations of the spins of both the transmitted electron and the
QD~\cite{Oreg}.

\subsection{The single-level Anderson model.}

Far from the charging resonances the interaction of the lead
electrons with the dot is described by the Kondo Hamiltonian
 \bq\label{HKondo} H_K= \sum_{iks} \eps_k c^{i\dagger}_{ks}
c^i_{ks}
 + \sum_{ij} [J^{ij}_0\hat{\vec \sigma}^{ij}_e{\vec S}_d
+V^{ij}_0\hat{n}^{ij}_e] \ .
 \ee
Here $i$ and $j$ denote left(L)
and right(R) leads, the operator $c^i_{ks}$ annihilates
the electron in lead $i$ with momentum $k$ and spin $s$. The
Pauli operators and the density of the conduction-electrons
on the impurity are given by
 \bq\label{density}
\hat{\vec \sigma}^{ij}_e=\sum_{k k's
s'}c^{i\dagger}_{ks} {\vec \sigma}_{ss'}c^j_{k's'} \ , \
\hat{n}^{ij}_e=\sum_{k k's}c^{i\dagger}_{ks} c^j_{k's} \ .
 \ee
Explicit formulas for $J_0$ and $V_0$ will be found below from the
tunnelling Hamiltonian describing the QD. We will consider a
time-reversal symmetric system. Then the matrices $J^{ij},V^{ij}$
are real and symmetric.

The interaction of the conduction electron with the spin of the
dot may be diagonalized by an orthogonal
transformation~\cite{Pustilnik} of $c_L,c_R$ into new
operators $c_u$ and $c_v$, described by the angle $\theta$, $\tan
2\theta={2J_0^{LR}}/{(J_0^{LL}-J_0^{RR})}$. This gives
 \bq\label{HKuv}
H_{\sigma S}= (J_0^u\hat{\vec \sigma}_u+
J_0^v\hat{\vec \sigma}_v){\vec S}_d \ .
 \ee
As long as $J_0^u,
J_0^v\ll 1$ the two couplings are renormalized independently
 \bq\label{RGuv}
1/J^{u,v}=1/J_0^{u,v} +
4\nu\ln\left( {T}/{\Gamma} \right).
 \ee
Here $\nu = \nu_L=\nu_R$ is the density
of states in the leads. In the case of
antiferromagnetic coupling ($J > 0$ ) this formula may be written
in the form ${4\nu}J=1/{\ln(T/T_K)}$ with $T_K$ being the
Kondo temperature. Crucial for the understanding of the phase
behavior is that in the leading order only the spin-dependent part
of the Hamiltonian (\ref{HKondo}) is renormalized, while the
scalar coupling remains unchanged,  $V=V_0=const$.

In the simplest case of only one level in the dot (Anderson
impurity model), only one mode $c=(t_Lc_L +t_Rc_R)/t$ is coupled
to the dot via the tunnelling matrix element
$t=\sqrt{|t_L|^2+|t_R|^2}$, while the second mode remains
completely decoupled. The bare values of the coupling constants in
the Kondo Hamiltonian (\ref{HKondo}) are now given by the second
order of perturbation theory
 \bq\label{bare}
\fr{t^2}{-\eps_d}=V_0+\fr{J_0}{2}  \ ; \
\fr{t^2}{-(U_{CB}+\eps_d)}=V_0-\fr{J_0}{2} .
 \ee
Here $\eps_d<0$ is the energy of the impurity level and $U_{CB}$
is the charging energy and the Fermi energy is $E_F=0$.

The non-spin-flip transmission  amplitudes for parallel and
antiparallel spins of the dot and the electron are
 \begin{eqnarray}\label{Aren}
&&A^d_{\uparrow\uparrow}=\fr{t_L^\dagger t_R}{-\eps_d}
\left(1
-\fr{i\Gamma}{-2\eps_d}\right) \ , \\
&&A^d_{\uparrow\downarrow} = \fr{t_L^\dagger
t_R}{-(U_{CB}+\eps_d)} \left(1
-\fr{i\Gamma}{-2(U_{CB}+\eps_d)}\right) \ , \nonumber
 \end{eqnarray}
where $\Gamma=2\pi t^2\nu$. The imaginary part of the amplitudes
here is formally of the fourth order in the tunnelling amplitudes
$t_{L,R}$. However, the calculation of the phase of transmission
amplitude requires only the $S$-matrix for non-spin-flip
scattering to second order in $t$. That is why the Kondo
correlations do not contribute to the phase in the leading order
and  Eq.~(\ref{Aren}) coincides with the expansion of the usual
Breit-Wigner formula. The Kondo effect appears in the real part of
the amplitudes at the order $\sim t^4$, which we will take into
account via the renormalization~(\ref{RGuv}). In order to find the
AB current one should rewrite the sum of the two amplitudes
$A^d_{\uparrow\uparrow}$ and $A^d_{\uparrow\downarrow}$ in terms
of scalar and magnetic couplings~(\ref{bare}) and then replace
$J_0$ by $J$~(\ref{RGuv}), which gives
 \bq\label{result}
G_{AB}\propto Re A^{r*} [ V_0-i\pi\nu (V_0^2+{J^2}/{4})] \ .
 \ee
This formula is the central result of this section. It should be
compared with a usual conductance of a Kondo quantum dot, obtained
by adding $|A_{\uparrow\uparrow}|^2$,
$|A_{\uparrow\downarrow}|^2$~(\ref{Aren}) and the corresponding
spin-flip contribution,
 \bq\label{Gusual}
G=2\fr{e^2}{h}\Gamma^2\fr{|t_L t_R|^2}{t^8} \left(
V_0^2+\fr{3}{2}J^2\right) .
 \ee
Remember that $V_0$ is not renormalized, while $J\sim
1/\ln(\Gamma/T_K)$. Eqs. (\ref{Gusual}) and (\ref{result}) are
justified only for $\nu J\ll 1$. This does not allow for the
quantitative description of the conductance in the unitary limit,
where $G\approx 2e^2/h$. On the other hand, the phase shift close to
$\pi /2$ develops at \bq \nu J_0\ll \nu J\sim\sqrt{\nu V_0}\ll 1.
\ee This gives the temperature scale, explaining the high
sensitivity of the transmission phase to the Kondo effect observed
in the experiment~\cite{Yang1}: $\ln(T/T_K)\sim
\sqrt{\ln(\Gamma/T_K)}$.

\begin{figure}[t]
\centering
\includegraphics[width=.8\textwidth]{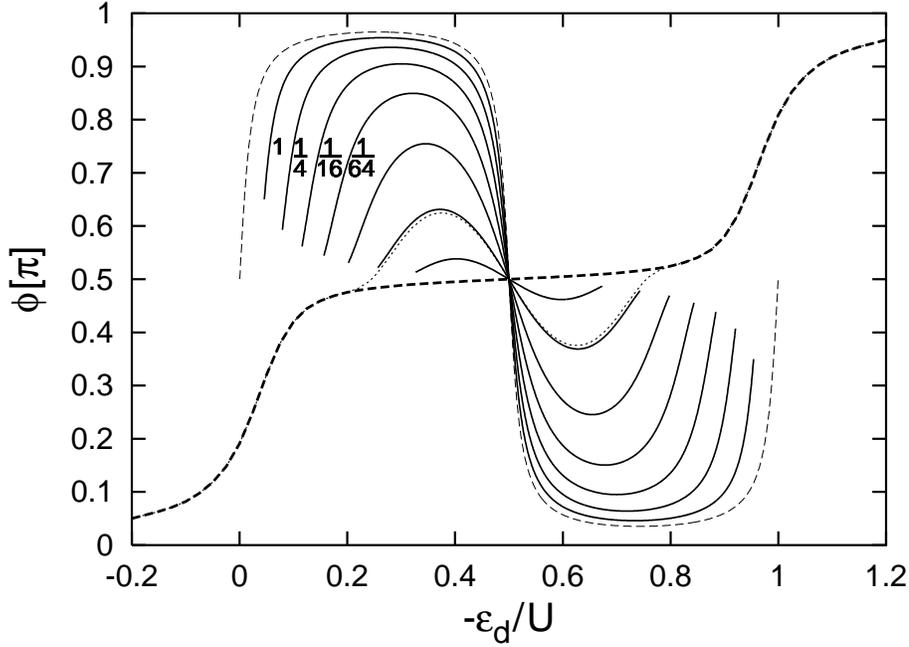}
\caption{ The phase $\phi$ as a function of the depth of the
impurity level $-\eps_d$ for $\Gamma=U/30$. Solid lines show the
calculated $\phi(\eps_d)$ for $T=\Gamma, \Gamma/4, ... ,\Gamma /4^6
$. Since the theory is valid only for $T\gg T_K$, the curves are
shown only for $T>2 T_K(\eps_d)$. The dotted line depicts
schematically the expected $\phi(\eps_d)$ for $T=\Gamma/1024$. The
thin dashed line shows the phase for the sum of two Breit-Wigner
resonances and the thick dashed line is the Bethe-ansatz solution
for $T\equiv 0$.}
\end{figure}

Fig.~11 shows the phase $\phi(\eps_d)$ found via
Eq.~(\ref{result}) for $\Gamma=U_{CB}/30$ and different
temperatures. Due to a node of $V_0$ at $\eps_d=-U/2$ the phase
equals exactly $\pi/2$ in the middle of the valley. The width of
the phase drop at higher temperatures ($T\sim \Gamma$) is
$\sim\Gamma$. The exact Bethe-ansatz solution~\cite{Bethe} for
$T\ll T_K$ predicts a broad plateau between resonances with
$\phi\approx \pi/2$~\cite{Delft}. However, due to a strong
dependence of $T_K$ on the position of the impurity level $\eps_d$
the phase changes from $\phi=0,\pi$ to $\phi=\pi/2$ in a rather
nonuniform way. For intermediate temperatures the phase first
develops a shoulder with $\phi\approx\pi/2$ ($T\ll T_K(\eps_d)$),
while in the center~($T\gg T_K(\eps_d)$) pronounced maximum and
minimum are formed to the left and right of the point
$\eps_d=-U_{CB}/2$. This structure may be seen on the experimental
figures of ref.~\cite{Yang1}, but appears  not to have been
discussed in the numerical calculation of ref.~\cite{Delft}. The
conductance (\ref{Gusual}) in the middle of the valley for the
same parameters of the dot as in Fig.~1 varies from $G=0.013
e^2/h$ for $J=J_0$ to $G=0.42 e^2/h$ for the lowest temperature
$T=\Gamma/4^6$. We see that the phase behavior is well developed
when the conductance is sufficiently below the unitary limit.

\section{Conclusions}

To conclude, we have considered the model for which upon
increasing the plunger gate voltage $V_g$, it is energetically
favorable to first populate in the dot the level strongly coupled
to the leads. At a somewhat larger $V_g$ a sharp jump occurs to a
state where the "next in line" narrow level $1$ becomes populated.
This jump occurs due to the appearance of $U_{CB}$ in some of the
denominators of the second-order energy shifts. It is thus a
genuine correlation effect and can not be faithfully addressed via
a Hartree-Fock-type approximation. The jump accounts for the sharp
decrease by $\sim \pi$ of the transmission phase. The similar
strengths of resonances seen in the experiment~\cite{Heiblum2} and
their large width are also clear within our mechanism. It appears
that this is the only available model which can explain these two
important features of the experiment. Even independently of that,
we believe that the level-occupation switching is interesting in
its own right. It can be further studied experimentally using a
combination of charge sensing and transport techniques. The
current transmission through such a QD resembles the behavior of
rare earth elements, whose chemical properties are determined not
by the electrons with highest energy, but by the "strongly
coupled" valence electrons. The overlapping of single-particle
resonances may take place also in the Kondo experiments in QD-s
\cite{Kondo1,Kondo2,Kondo3,Kondo4,Kondo5}, where in order to
increase the Kondo temperature the dot is usually sufficiently
opened. Hopefully the unusual effects observed in some of these
experiments may be also explained within our approach.

Our mechanism of charging of the QD requires the existence of the
broad level with $\Gamma\gg\Delta$. The simple way to examine the
relevance of our theory for the explanation of the experiment of
ref.~\cite{Heiblum2} will be to close the dot sufficiently in order
to have $\Gamma\ll\Delta$ for all levels. In this case the phase
still increases by $\pi$ at any resonance, but the correlation
between peaks will disappear. (More precisely the pairs of peaks
corresponding to adding of electrons with opposite spins onto one
and the same level are still correlated, but the correlation between
pairs should disappear.) Moreover, within our mechanism, a series of
$\sim(\Gamma/\Delta)\ln (U_{CB}/\Gamma)$ strong charging peaks in
the conductance should have the same height. This ``coupling
dependent'' correlation of the peak heights seems also easy to
measure.

Next, we discussed the effects of spin. Here the most interesting
predictions are the two-electron occupation switchings in the valley
and sharp features inside the charging resonances. For the first
time here we considered some fine structures of the population
switching and their signatures in the signal of the charge detector.

In the last section,  the extra-strong sensitivity of the
transmission phase through a quantum dot to the Kondo correlations
was explained. The nontrivial behavior of the phase develops at
temperatures large compared to $T_K$ and may be found analytically
by means of a simple renormalization group calculation. The phase
reflects the developing Kondo correlations in the regime where,
although the spin of the impurity is not screened, the running Kondo
coupling constant exceeds parametrically its bare value. New
features of $\phi$, which were not noticed in existing numerical
simulations were found for the Anderson impurity model.

Finally, we mention that the results of a recent experiment on the
phase behavior in very small dots with not-too-large electron
numbers~\cite{Michal} would fit very naturally with models that
necessitate relatively large level-width to level-separation
ratios for the universal phase correlations. This is because the
level-separation obviously increases with decreasing dot sizes. In
fact, for the small dots the universal behavior gives way to more
random "mesoscopic" one. The universal to mesoscopic crossover was
studied in Refs.~\cite{KarrFano,KarrNJP}, with emphasis on the
relevance of Fano physics to this problem. Numerical studies of
the interplay of interference, interactions and level-occupation
switching have been presented in Ref.~\cite{FanoPS}.  Systematic
experimental studies invoking both the size and the opening of the
dot as well as charge sensing together with transport, could shed
more light on these fascinating phenomena.

 \ack
Discussions with N.~Andrei, E.~Buks, D.~Esteve, A.~Georges,
L.~I.~Glazman,  M.~Heiblum, Y.~Gefen, K.~Kikoin, I.~Lerner,
Y.~Levinson, Y.~Oreg, M.~Pustilnik, M.~Schechter, V.~V.~Sokolov,
D.~Sprinzak, H.~A.~Weidenmuller and A.~Yacoby are greatly
appreciated. This work was supported by the SFB TR 12, by the German
Federal Ministry of Education and Research (BMBF) within the
framework of the German-Israeli Project Cooperation (DIP) and by the
Israel Science Foundation (grant No. 1566/04). PGS's visit at
Weizmann was supported by the EU - Transnational Access program, EU
project RITA-CT-2003-506095.


\begin{thebibliography}{99}
\bibitem{endnote} The two-terminal setup of the early experiment
Ref.~\cite{Heiblum1} does not allow to measure the phase of
transmission amplitude through a QD. In order to be able to
measure the transmission phase, the interferometer must be opened
up to allow electrons to escape from it. This converts the problem
to an n-terminal one where at least $n>2$ \cite{Weidenmuller}. The
nontrivial requirements from the "opening" process to produce the
correct phase have been discussed in ref \cite{EAIH}.
\end{thebibliography}
\end{document}